\documentclass[a4paper]{mcs13}

\usepackage{tabularx} 
\usepackage{graphicx} 
\usepackage[intlimits]{amsmath}
\usepackage{amssymb} 
\usepackage{times}
\usepackage{soul} 
\usepackage[square,numbers,sort&compress]{natbib} 
\usepackage[labelfont=bf,font=normalsize]{caption}  
\usepackage{epstopdf} 
\usepackage{hyperref}
\usepackage{multirow}
\usepackage[version=4]{mhchem}
\hypersetup{
    colorlinks=true,
    citecolor = blue,
    linkcolor=blue,
    urlcolor = blue
}

\begin{document}

\title{Machine learning enhanced multi-particle tracking in solid fuel combustion}

\authors{H. Chen*, Y. Li*, B. Böhm* and T. Li*}

\corrauthoremail{tao.li@rsm.tu-darmstadt.de}

\address{*Technical University of Darmstadt, Department of Mechanical Engineering, Reactive Flows and Diagnostics, Otto-Berndt Straße 3, 64287 Darmstadt, Germany}

\begin{abstract}
Particle velocimetry is essential in solid fuel combustion studies, however, the accurate detection and tracking of particles in high Particle Number Density (PND) combustion scenario remain challenging.~The current study advances the machine-learning approaches for precise velocity measurements of solid particles.~For this, laser imaging experiments were performed for high-volatile bituminous coal particles burning in a laminar flow reactor.~Particle positions were imaged using time-resolved Mie scattering.~Various detection methods, including conventional blob detection and Machine Learning (ML) based You Only Look Once (YOLO) and Realtime Detection Transformer (RT-DETR) were employed and bench marked.~Particle tracking was performed using the Simple Online Realtime Tracking (SORT) algorithm.~The results demonstrated the capability of machine learning models trained on low-PND data for prediction of high-PND data.~Slicing Aided Hyper Inference (SAHI) algorithm is important for the better performance of the used models.~By evaluating the velocity statistics, it is found that the mean particle velocity decreases with increasing PND, primarily due to stronger particle interactions.~The particle dynamics are closely related to the position of combustion zone observed in the previous study.~Thus, PND is considered as the dominant factor for the particle group combustion behavior of high-volatile solid fuels.
\end{abstract}

\section{Introduction}

Understanding the solid fuel (SF) combustion process is crucial for the advancement of various industrial applications, including electricity production, propulsion systems, and environmental protection.~The fundamental of SF combustion, which is a complex chemical reaction involving fuel and oxidizer in multiple phases, relies on the understanding of the physical, chemical and other interactive processes in flows.~Obtaining such knowledge require accurate experimental data, such as particle dynamics during the combustion process, which is non-trivial to collect.

Particle velocity is an essential parameter closely related to combustion behavior \cite{Li.2023}.~Balusamy et al.~\cite{Balusamy.2013.Exp.Fluids} reported that larger particles exhibited lower velocity, which was measured by both particle image velocimetry (PIV) and Laser Doppler velocimetry (LDV).~Attili et al.~\cite{Attili.2021} noted that gas-phase ignition delay time decreased with increasing particle slip velocity from simulations.~Li et al.~\cite{Li.2020.Proc.Combust.Inst.} measured particle velocity using PIV and observed that particles near the boundary of a particle cloud exhibited higher velocities compared to those at the center.

In literature, different techniques have been employed to measure particle velocity during combustion.~The LDV was used to measure the velocities of SF particles~\cite{Balusamy.2013.Exp.Fluids} by analyzing the Doppler shift of the laser light scattered by particles.~Yao et al.~\cite{Yao.2019} employed high-speed Digital In-line Holography (DIH) to image burning coal particles and subsequently calculated particle velocity using the algorithm described in~\cite{Yingchun.2014}.~PIV is another widely used method due to its simple setup and robustness.~Particle motion in fluids is captured by taking two sequential images in a short time interval and evaluating the displacement of particle patterns based on cross-correlation \cite{Raffel.2018}.~To resolve the particles, we adopted Particle Tracking Velocimetry (PTV), a technique that measures the velocity of individual particles by tracking their positions over time.~PTV has been utilized in previous research~\cite{Ahmadi.2019.Int.J.Multiph.Flow,Yingchun.2014} to measure the velocity of glass beads and particles in digital holography.~However, according to Dracos et al.~\cite{Dracos.1996}, PTV is not suitable for high PND scenarios due to the difficulty in distinguishing individual particles.~To address this issue, ML-based detection methods are proposed to enhance the accuracy of particle velocity calculation.

Building upon the previous work by Li et al.~\cite{Li.2022.Proc.Combust.Inst.,Li.2023.Fuel}, this study aims to further explore particle motion in high PND scenarios using PTV based on machine learning detection methods.~The primary objectives are to assess the feasibility of ML techniques to train models on simpler data and subsequently apply them to more complex scenarios, thereby enhancing the model generalization and overall accuracy of particle velocity calculation.~The models are validated on tracking a single-particle trajectory.~The velocity statistics are evaluated, showing clear dependence on the particle-particle interaction.~In the following, a concise overview of the experimental setup and the data set used in this study are firstly introduced.~Subsequently, traditional blob detection algorithm, ML methodologies, and the performance assessment of implemented detection and tracking methods are elaborated.~The outcomes of these various approaches are presented and discussed. 

\section{Experimental data set}\label{Sec:Exp}
\subsection{Experimental setup}

\begin{figure}[!ht]
	\centering
	\includegraphics[width=140mm]{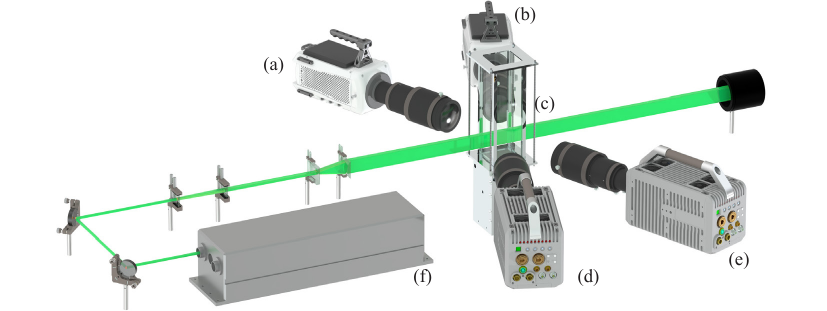}
	\caption{Experiment setup~(a)(b) Phantom v711 cameras~(c) Laminar flow reactor~(d)(e) Photron SA-X2 cameras~(f) Diode-pumped laser}
	\label{fig:Fig_Setup_GPC}
\end{figure}

In order to study the combustion process of coal particles, optical experiments focusing on particle groups burning in laminar flow reactor were performed.~Figure~\ref{fig:Fig_Setup_GPC} illustrates the experimental setup of the optical measurement.~The detailed experimental setup was reported previously~\cite{Li.2022.Proc.Combust.Inst.} and is briefly introduced here.~A laminar flow reactor (LFR), Figure~\ref{fig:Fig_Setup_GPC} (c), was used to provide solid fuel particles into a gas atmosphere with well-defined temperature, and species concentrations.~A fully premixed flat flame was stabilized above the ceramic honeycomb structure.~By adjusting the flow rate of~\ce{CH4},~\ce{O2} and~\ce{N2}, the desired post-combustion gas with homogeneous temperature and flow profile can be obtained \cite{Li.2021.Fuel}.~These gas conditions were denoted as A10/A20/A30/A40 representing the 10\%, 20\%, 30\% and 40\% volume fraction of oxygen in the gas atmosphere, while A20 is focused in the current work.~High-volatile bituminous (hvb) coal particles were injected into the burner through a capillary tube with a diameter of 3\,mm by carrier flow with same gas composition and velocity as the flat flame inlet gas.

Particles were visualized by the high-speed Mie-scattering measurements at 10\,kHz, as reported in \cite{Li.2022.Proc.Combust.Inst.}.~A diode-pumped single-head ND:YAG laser~(Innoslab, Edgewave), as shown in Figure~\ref{fig:Fig_Setup_GPC} (f), with a pulse energy of 2.5\,mJ at 532\,nm.~To cover the entire particle region, the laser beam thickness was expanded to approximately 8\,mm.~For a later employment of tomographic PTV, four CMOS cameras were used that are equipped with with macro lens~(Sigma, $f$\,=\,180\,mm, $f$/32, DOF \textgreater 10\,mm).~A pair of Photron SA-X2 cameras~((d) and (e) in Figure~\ref{fig:Fig_Setup_GPC}) were set up with $32^\circ$ to the laser beam.~On the other side, another pair of Phantom v711 cameras ((a) and (b) in Figure~\ref{fig:Fig_Setup_GPC}) were set with a angle of $16^\circ$.~The exposure time was set to 5\,$\mu$s to reduce the interference signal from the luminescent flame.~The field of view (FOV) of the Mie-scattering covered an area of 37.7 (height) $\times$ 21.6 (width) $\mathrm{mm^2}$ with a pixel resolution of $36.8\,\mu$m/pixel.~The extensive FOV enabled by Mie-scattering facilitates the acquisition of the particle position and velocity across the full duration of the combustion process.~In this paper, only data from the (d) camera was used to develop the algorithm for particle detection and tracking.~Tomographic PTV using the data from the other cameras will be investigated in the future.

\subsection{Experimental data}
\label{subsec:expdata}

In this experiment, hvb coal particles with a mean diameter of 125\,\textmu m were investigated.~The experimental data were divided into four cases that are shown in Table.~\ref{tab:expdata}.~The cases are defined as P$\alpha$A$\beta$, where P represents the particle combustion, $\alpha$ denotes the particle number ($N_\mathrm{par}$), and A represents the gas condition, and $\beta$ indicates the oxygen mole fraction.~Detailed calculation of $N_\mathrm{par}$ and corresponding PND is elaborated in the result section.

\begin{table}[!ht]
\caption{Experiment data sets}
\label{tab:expdata}
\centering
\small
\begin{tabular}{ccccc}

\hline
Cases   & Particle & Gas condition & $N_\mathrm{par}$ & PND ($\mathrm{mm}^{-3}$) \\ \hline
P040A20 & hvb Coal     & A20           & \textless{}40    & \textless{}0.18        \\
P060A20 & hvb Coal     & A20           & (40,60{]}        & (0.18,0.22{]}          \\
P080A20 & hvb Coal     & A20           & (60,80{]}        & (0.22,0.27{]}          \\
P100A20 & hvb Coal     & A20           & \textgreater{}80       & \textgreater{}0.27     \\ \hline
\end{tabular}
\end{table}

\section{Detection and tracking approaches}
\label{Sec:methods}

This section outlines the methodologies employed for particle position detection and tracking.~The detection methods include both traditional blob detection techniques and advanced machine learning approaches, such as YOLO~\cite{GitHub.20241112} and RT-DETR~\cite{Zhao.2023}.~For particle tracking, the SORT~\cite{Bewley.2016} algorithm is utilized.

\subsection{Blob detection}
\label{subsec:bolbdetection}

Blob detection techniques are designed to identify circular or blob-like structures in images, making them suitable for SF particle analysis.~The common approach for edge detection in blob detection is the Laplacian operator ($\nabla^2$), which measures the concentration or dispersion at a given point within a scalar field by computing the gradient field's divergence.~A negative result indicates a local maximum (potentially noise or an edge), while zero indicates a flat region.~To minimize noise, the image is smoothed using the Gaussian operator ($G(x,y,\sigma)$).~Thus, the Laplacian of Gaussian (LoG), which combines the Gaussian and Laplacian operators, is suitable for blob detection.~The Gaussian and LoG operators are given as follows~\cite{Lindeberg.2011}:

\begin{equation}
	G(x,y,\sigma)=\frac{1}{2\pi\sigma^2}e^{-\frac{x^2+y^2}{2\sigma^2}}
\end{equation}
\begin{equation}
	LoG(x,y,\sigma)=\nabla^2G(x,y,\sigma)=\frac{x^2+y^2-2\sigma^2}{\sigma^4}G(x,y,\sigma)=\frac{x^2+y^2-2\sigma^2}{2\pi\sigma^6}e^{-\frac{x^2+y^2}{2\sigma^2}}
\end{equation}

When $\sigma = \frac{r}{\sqrt{2}}$, the LoG operator maximizes its response on circular blobs with $r$ standing for the radius of the blob, allowing detection of varying blob sizes by adjusting $\sigma$.~As $\sigma$ increases, the LoG response becomes smoother and weaker.~But this variation will influence the $\sigma$ selection since the maximum response occurs at the finest scale and the minimum at the coarsest.~To compensate this response decrease at larger $\sigma$, a factor $\sigma^2$ is applied, resulting in the normalized LoG operator:

\begin{equation}
	LoG(x,y,\sigma)_{norm}=\sigma^2\nabla^2G(x,y,\sigma)=\frac{x^2+y^2-xy\sigma^2}{2\pi\sigma^6}e^{-\frac{x^2+y^2}{2\sigma^2}}
\end{equation}

Although it is feasible to detect the blob by locating the maximum response of $LoG_{\text{norm}}$, the calculation is time-intensive.~Hence, the Difference of Gaussian (DoG) operator is introduced as an efficient approximation.~Defined as follows, DoG approximates LoG by computing the difference between two Gaussian operators~\cite{Lowe.2004}.~When considering the difference of the nearby scale at $k\sigma$ and $\sigma$, the DoG operator approximates the normalized LoG as described in the following equation:



\begin{equation}
	DoG=G(x,y,k\sigma)-G(x,y,\sigma) \approx (k-1)LoG_{norm}
\end{equation}

Here, $k$ represents the standard deviation ratio used to compute the DoG.~The DoG operator is computationally efficient while retaining the capability to detect blobs across various scales.

\subsection{Machine learning detection}
\label{subsec:mldetection}

Although DoG has significant performance on particle detection, it still requires complex pre-processing for different combustion cases, which is time-consuming.~Inspired by the development of recently developed ML-based object detection models, the implementation of YOLOv8 and RT-DETR is anticipated to be trained on a small scale of low-PND case data and then applied to high-PND case data.~This section elaborates the principle of the used models, as well as the performance evaluation methods.

The objective of ML object detection is to find and classify the objects in images by providing the bounding boxes and the class labels.~The primary approaches to achieve object detection include two-stage methods and one-stage methods.~Examples of the two-stage methods include R-CNN~\cite{OShea.20151127}, Res-Net~\cite{He.20151211}, and fast R-CNN~\cite{Girshick.2015430}.~In the first stage, a Region Proposal Network (RPN)~\cite{Fang.2017} generates a set of candidate object regions, also known as \textit{region proposals}, within the image.~This step focuses on identifying potential areas that may contain objects, significantly reducing the search space.~In the second stage, these region proposals are further processed to classify objects and refine bounding box predictions.~The two-stage methods possesses higher accuracy but requires more computation resources.~On the contrast, the one-stage approach eliminates the \textit{region proposal} step and directly performs classification and localization in a single pass over the image.~Models such as YOLO , as its name indicates, is an representative model using this approach.~By predicting bounding boxes and object classes for the entire image at once, one-step models achieve significantly faster detection speeds.~As the development of the one-stage models, a new model RT-DETR was developed based on transformer architecture.~Considering the one-stage models show significantly better performance on accuracy and speed, both YOLO and RT-DETR are adopted in this study for particle detection.

\subsubsection{YOLO}
\label{subsec:yolo}

After years of iteration, the latest version of YOLO has already presented a fantastic performance on object detection purpose.~This is contributed by its innovated design, which works in following steps.~It firstly divides the image into a grid of certain size.~Each grid cell is responsible for detecting objects whose centers fall within the cell.~Then in each grid cell, YOLO predicts a certain number of bounding boxes consisting of coordinates and a confidence score.~The coordinates include the object center as well as the width and height of the bounding box (both relative to the whole image).~After that, the cell predicts a probability distribution over predefined object classes within the cell and the assigns the highest class probability to the bounding box with highest confidence score.~Finally, it will eliminate the redundant bounding boxes by applying the Non-Maximum Suppression (NMS).

\begin{figure}
	\centering
	\includegraphics[width=140mm]{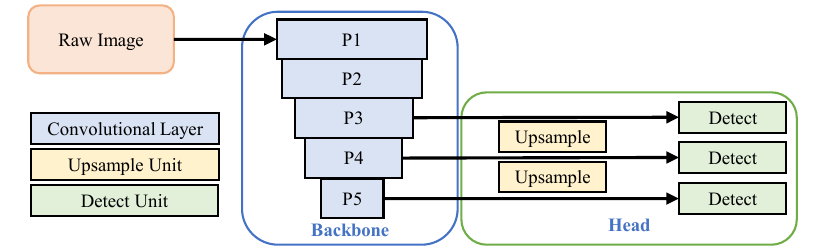}
	\caption{The backbone and head structure of YOLOv8 detection model}
	\label{fig:YOLO8}
\end{figure}

The Fig.\ref{fig:YOLO8} briefly illustrates the architecture of YOLOv8 that has been implemented in this work.~It consists of a backbone to extract features at different scales and 3 detection heads to make the prediction.~The backbone, which includes several convolutional layers and special blocks, is able to reduce the spatial dimensions while increasing the channels of the feature maps.~Then, the head part up-samples and concatenates the feature maps from different scales and transmits them to the detection heads.~The detection head will finally generate the possible bounding boxes.

\subsubsection{RT-DETR}
\label{subsec:RT-DETR}

The second machine learning model adopted for the current experimental data is RT-DETR.~The key feature of the RT-DETR include the end-to-end object detection architecture, transformer-based attention mechanism, and use of set-based loss for direct object detection~\cite{Zhao.2023}.~These features allow DETR to simplify the object detection pipeline and achieve high performance.

\begin{figure}
	\centering
	\includegraphics[width=140mm]{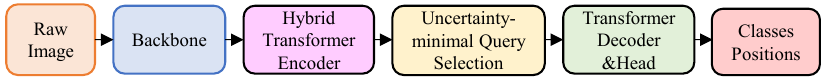}
	\caption{The architecture of the RT-DETR model}
	\label{fig:RT-DETR}
\end{figure}

Comparing to the YOLO model, the prediction head of the architecture is replaced by the transformer-based structure as shown in Fig.~\ref{fig:RT-DETR}.~The transformer encoder-decoder structure is used to capture global context across the entire image.~The self-attention mechanism allows the model to understand relationships between different parts of the image, enhancing its ability to recognize and localize objects, even those that are small or partially occluded.~The transformer decoder introduces object queries, which act as learnable embeddings that interact with the feature map to detect specific objects.~Each object query learns to focus on a unique part of the image, enabling efficient object detection without relying on anchor boxes.~The set-based loss function assigns a one-to-one match between predicted and ground truth objects.~This Hungarian matching algorithm ensures that each object is detected only once, eliminating the need for NMS to associate the bounding boxes with the grid cell.~The set-based loss encourages the model to focus on unique object instances, reducing duplicate detection.~The performance of RT-DETR will be further discussed in the result section.

\subsection{Detection performance evaluation}
\label{subsec:detectperforman}

To compare the performance of traditional blob detection method and the machine learning enabled detection, an evaluation method based on confusion matrix was adopted.~The employed metrics were able to statistically analyse the agreement between detected bounding boxes and the ground truth bounding boxes.~To evaluate the prediction performance, the ground truths were manually labeled using the open source detection labeling tool \textit{LabelImg}~\cite{GitHub.20241112b}.~The similarity of the two bounding boxes are quantified by Intersection over Union (IoU).~In the detection field, the IoU is equal to the overlap area between the bounding box of detection $B_{det}$ and the bounding box of ground-truth $B_{gt}$ over the the area of their union ($\mathrm{IoU} =\frac{area(B_{det}\cap B_{gt})}{area(B_{det}\cup B_{gt})}$).


When IoU$=1$, it indicates a perfect match between the two bounding boxes, while IoU$=0$ signifies no overlap.~Thus, a higher IoU value corresponds to better detection accuracy.~An IoU threshold is set to assess detection quality: values above the threshold are considered \textit{true}, and those below are \textit{false}.

\begin{figure}[h!]
	\centering
	\includegraphics[width=70mm]{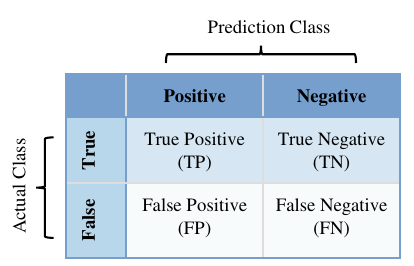}
	\caption{The structure of confuse matrix classes.~The four classes are shown in the Figure. }
	\label{fig:matrix}
\end{figure}

To evaluate the detection performance, the confusion matrix is used.~Four detection classes are defined: True Positive (TP), True Negative (TN), False Positive (FP), and False Negative (FN), as shown in Figure~\ref{fig:matrix}.~'True' and 'False' indicate the correctness of the model's detection as described before, while 'Positive' and 'Negative' refer to the existence of the ground-truth detection.

When evaluating the performance of detection methods, it is essential to consider not only the accuracy but also the coverage of the detections.~An ideal model should detect all possible targets with high precision.~Therefore, the concepts of precision and recall are introduced to assess the model's performance.~Precision indicates the model's ability to correctly identify objects, represented by the percentage of true positive detections over all detections.~Recall reflects the model's capability to find all objects in the dataset, represented by the percentage of true positive detections over all ground-truth instances.~These parameters are derived from the confusion matrix as follows:

\begin{equation}
	\mathrm{Precision} =\frac{\rm{N}_{TP}}{\rm{N}_{TP}+\rm{N}_{FP}}=\frac{\rm{N}_{TP}}{\rm{N}_{det}} 
	\label{eq:Precision}
\end{equation}
\begin{equation}
	\mathrm{Recall} =\frac{\rm{N}_{TP}}{\rm{N}_{TP}+\rm{N}_{FN}}=\frac{\rm{N}_{TP}}{\rm{N}_{gt}} 
	\label{eq:Recall}
\end{equation}

\subsection{The tracking method}
\label{subsec:trackingmethod}

The SORT algorithm is a simple but effective method for Multiple Objects Tracking (MOT).~SORT is designed to track objects across frames in real-time, making it also suitable for particle tracking of the current data.~The algorithm is based on the principles of data association and motion prediction, using a combination of the Kalman Filter, Hungarian algorithm, and IOU-based bounding box matching to track objects efficiently~\cite{Bewley.2016}.

The SORT algorithm leverages a Kalman Filter to predict the subsequent position of each tracked object, relying on the object's prior state for this estimation.~The Kalman Filter effectively estimates both position and velocity, allowing for adjustments that accommodate minor variations in motion and account for noise between consecutive frames.~For data association, SORT employs the Hungarian algorithm, which optimally matches predicted object locations with new detections by minimizing the distance between them, typically quantified through Euclidean distance in the image plane.~Object detection matches in SORT are further refined using the IoU threshold.~A detection is assigned to an existing tracker only if the IoU exceeds a certain threshold (0.3 used in this experiment), ensuring that only closely aligned objects are paired.~In instances where new detections do not match any existing tracker, new trackers are initialized.~Conversely, trackers that fail to receive updates, or matching detections, within a predefined frame interval are removed, effectively filtering out noise and managing lost objects in the tracking process.


\section{Results and discussions}\label{Sec:results}

\subsection{Particle Number Density}
\label{subsec:pnd}

\begin{figure}[h!]
	\centering
	\includegraphics[width=140mm]{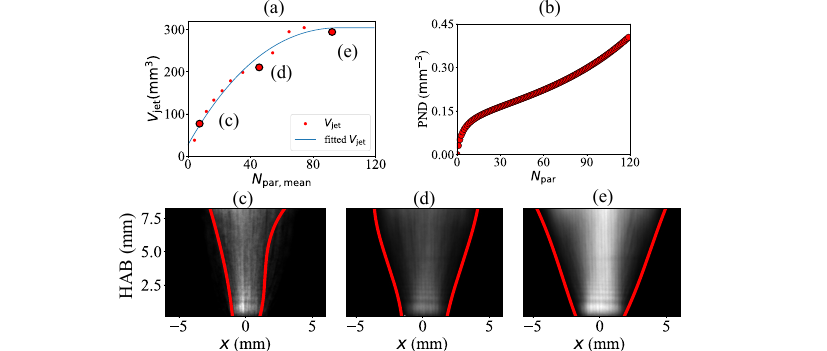}
	\caption{(a) The relationship of reference volume and $N_\mathrm{par}$.~The fitted volume is also shown.~(b) The PND and $N_\mathrm{par}$ ratio calculated using the fitted reference volume.~(c)(d)(e) The binary images of the particle jet of three different $N_\mathrm{par}$ groups with fitted particle jet boundary.}
	\label{fig:V_jet}
\end{figure}

The particle number density is an important parameter to assess particle-particle interaction, which is defined as PND=$N_\textrm{par}$/$V_{\textrm{jet}}$~\cite{Li.2022.Proc.Combust.Inst.}.~The reference volume $V_{\textrm{jet}}$ of the particle jet was derived using particle images across all cases.~These images were divided into 12 subgroups based on $N_\textrm{par}$ calculated within a HAB range of 0\,–\,7.8\,mm, where no soot particles form.~A threshold was applied to the averaged binary image of each group to determine the boundary of the particle jet.~This boundary was then fitted using a polynomial fitting function.~Subsequently, the reference volume was calculated based on this fitted boundary by assuming 3D symmetry.~The correlation between the reference volume and particle number is illustrated in Fig.~\ref{fig:V_jet} (a).~The particle jet trends to expand with increasing $N_\mathrm{par}$ (as shwon in Figure.\ref{fig:V_jet} (c), (d) and (e)) and remain unchanged after $N_\mathrm{par}$ reaches around 100, as indicated in Figure.~\ref{fig:V_jet} (a).~The curve was fitted and further used to calculate PND.~The relationship between PND and $N_\mathrm{par}$ is depicted in Fig.~\ref{fig:V_jet}(b) and is further used for velocity statistic analysis.~Compared to our previous work \cite{Li.2022.Proc.Combust.Inst.}, the particle reference volume shows high consistency, while PND and $N_\mathrm{par}$ are lower in this study.~This is mainly due to the fact that Mie scattering has a lower pixel resolution (36.8\,\textmu m/pixel) compared to the previous diffuse-backlight illumination measurements (9.2\,\textmu m/pixel).~Thus, potential line-of-sight overlapping of particles underestimate the particle number and PND.

\subsection{Particle detection}
\label{subsec:multiparticledetection}

\begin{figure}
	\centering
	\includegraphics[width=140mm]{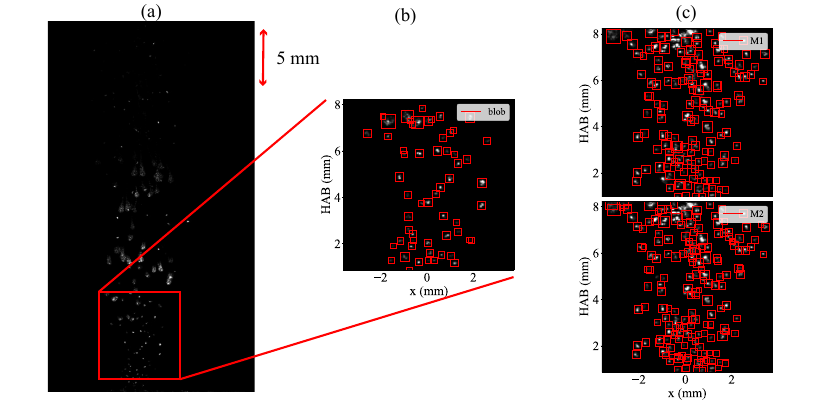}
	\caption{(a) The raw full scale image from P040A20 cases.~(b) Cropped and upscaled image with bounding boxes detected by blob detection for training.~(c) The bounding boxes detected by M1 and M2 of cropped and upscaled images from P100A20 cases. }
	\label{fig:training}
\end{figure}

One objective of our training plan is to assess the feasibility of training the model on simpler cases with less particles and subsequently applying it to complex cases with more particles.~As shown in Table~\ref{tab:training}, two models are trained, denoted as M1 and M2.~Firstly, both M1 and M2 were trained based on YOLOv8 and RT-DETR, respectively, including 80 images from the data P040A20, in which the particle number $N_\mathrm{par}$ is below 40.~The training data were cropped (200 $\times$ 200 pixel) from the raw image to emphasize the features of the particles, as shown in Fig.~\ref{fig:training} (a) and (b).~The two models all use blob detection as ground truth and were trained 300 iterations.~To evaluate the models' performance on complex cases, the trained models were validated on the cropped and up-scaled images from data set P100A20 with $N_\mathrm{par}$ \textgreater 80, as illustrated in Fig.~\ref{fig:training} (c).~The ground truth of test images was based on blob detection for a quick evaluation.~Both M1 and M2 exhibit a precision score higher than 75\% and recall higher than 80\%, indicating a well performance on particle detection.~These findings suggest that training with a low-PND data set still yield effective performance on high-PND cases.

Although both M1 and M2 exhibit high precision score, focusing on the bounding boxes of M1 and M2, as shown in Fig.~\ref{fig:training} (c), the M2 model trends to over predict ( produce more predictions on one particle) when particles form soot.~This may be caused by its transformer-based architecture, which lacks a NMS mechanism to filter out redundant detections.~The performance of M1 and M2 on full-scale image were further investigated.

\begin{table}[h!]
\small
\centering
	\caption{Training and Prediction Cases}
	\label{tab:training}

\begin{tabular}{cccccc|cccc}
\hline
\multicolumn{6}{c|}{Training}                                    & \multicolumn{4}{c}{Validation}        \\ \hline
Model & Data Set & Image number & Ground truth & epoch & Model   & Precision & Recall & IoU & Confidence \\ \hline
M1    & P040A20  & 80           & Blob         & 300   & YOLOv8  &     0.801      &    0.82    &   0.4  &    0.5        \\ \hline
M2    & P040A20  & 80           & Blob         & 300   & RT-DETR &     0.779     &    0.80    &   0.4  &    0.5        \\ \hline
\end{tabular}

\end{table}


\begin{figure}[h!]
	\centering
	\includegraphics[width=140mm]{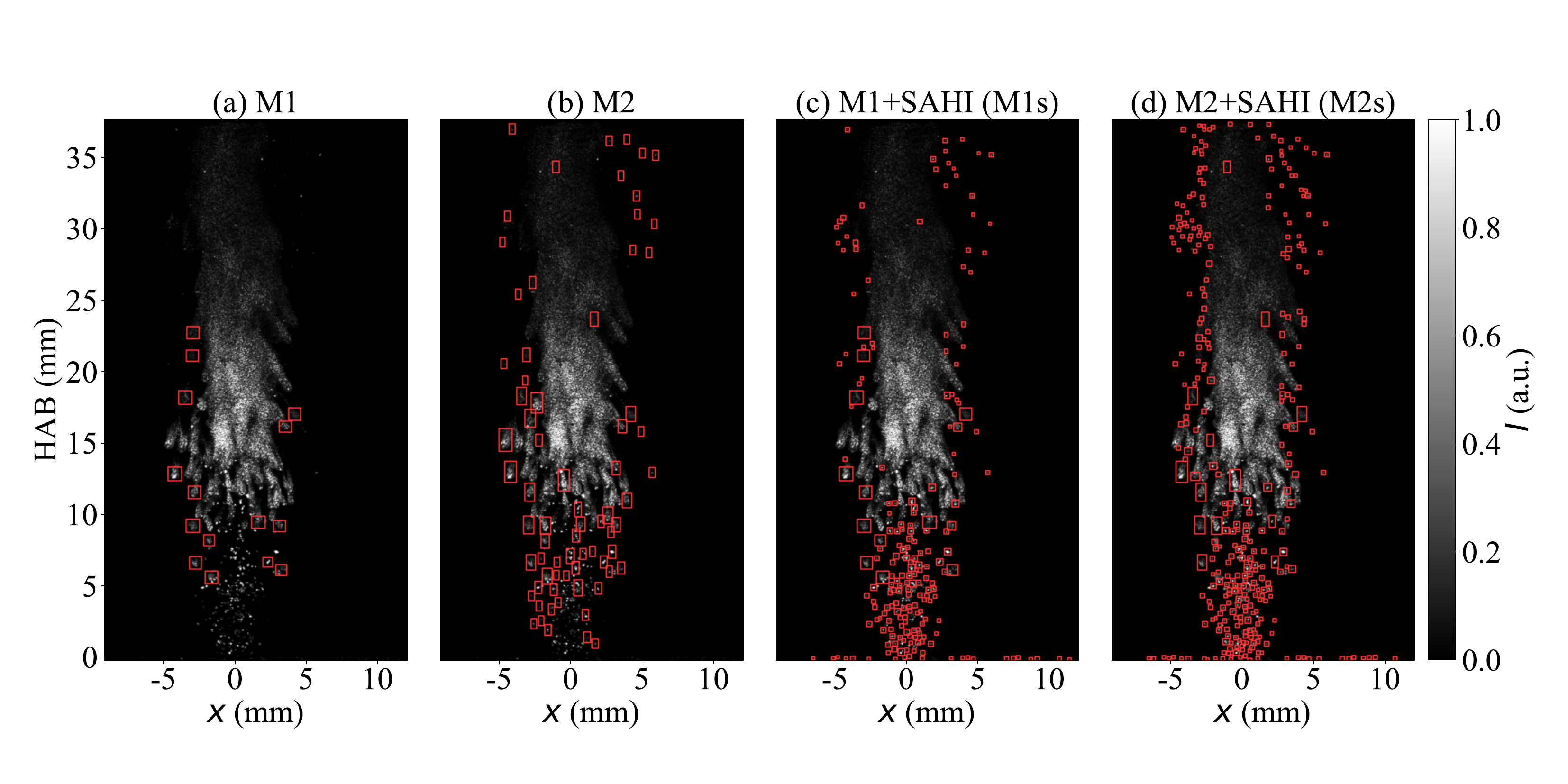}
	\caption{The bounding boxes detected by (a) M1, (b) M2, (c) M1 combined with SAHI (M1s), and (d) M2 combined with SAHI (M2s) in full scale image of P080A20 case.}
	\label{fig:det_all}
\end{figure}

When applied to full-scale images, both M1 and M2 exhibit sub-optimal performance as shown in Fig.~\ref{fig:det_all} (a) and (b).~This is due to the images being down-scaled to 640\,$\times$\,640 pixels when direct performing the detection model, which results in a loss of detail in the particle features.~To address this, the SAHI method is introduced, a strategy designed to improve detection of small or densely packed objects within large images.~SAHI employs a \textit{sliding strategy} that systematically divides large images into overlapping smaller patches, which are independently processed by M1 and M2.~Post-processing is then applied to merge results, using NMS to remove duplicate detections along slice boundaries, yielding cohesive detection results for the entire image.~As illustrated in Fig.~\ref{fig:det_all} (c) and (d), the SAHI method significantly improve M1 and M2 performance on full-scale images, particularly for particles with low contrast against the background.~However, similar to the results with cropped and up-scaled images, M2 combined with SAHI (M2s) continue to exhibit over-detection issues for particles with soot.~Therefore, M1 combined with the SAHI method (M1s) is selected as the final approach for particle tracking.

\subsection{Particle tracking}
\label{Sec:tracking}

\begin{figure}[htbp]
	\centering
	\includegraphics[width=140mm]{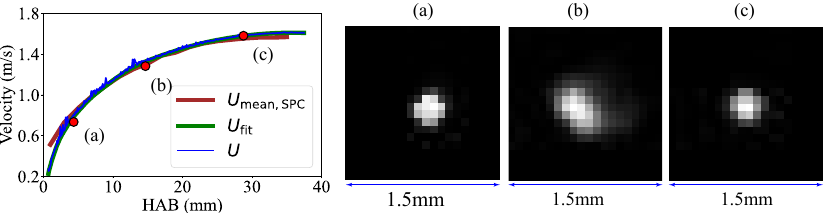}
	\caption{The velocity of a tracked particle in case P040A20 with detected ROI images of the particle.~The average velocity of particles from previous work~\cite{Li.2024} and the fitted velocity are presented for a comparison.}
	\label{fig:singletracking}
\end{figure}

After the detection, the SORT method is applied to acquire the particle velocity data.~In Fig.~\ref{fig:singletracking}, the tracking results for an individual particle in the P040A20 case, based on M1s, are presented.~The particle velocities are calculated using the Five-Point Difference Method based on the tracked particle center over time.~The left diagram in Fig.~\ref{fig:singletracking} illustrates the velocity changes of particles with HAB.~Due to the resolution limitations of Mie-scattering, particles appear about 10 pixels in diameter.~Consequently, even a positional error of one pixel can introduce significant noise in the velocity analysis.~To mitigate this, a polynomial fitting algorithm was applied to the velocities obtained from SORT tracking results, providing a more accurate velocity analysis.~The over all velocity is a little bit lower than the average velocity from previous work~\cite{Li.2024} ($U_{\textrm{mean,SPC}}$).~The right side of Figure~\ref{fig:singletracking} displays cropped particle images at various HABs.~Videos of tracking results with both cropped and full scales, provided in the supplementary materials, demonstrate the robust tracking capability of the SORT method.


\begin{figure}[h!]
	\centering
	\includegraphics[width=70mm]{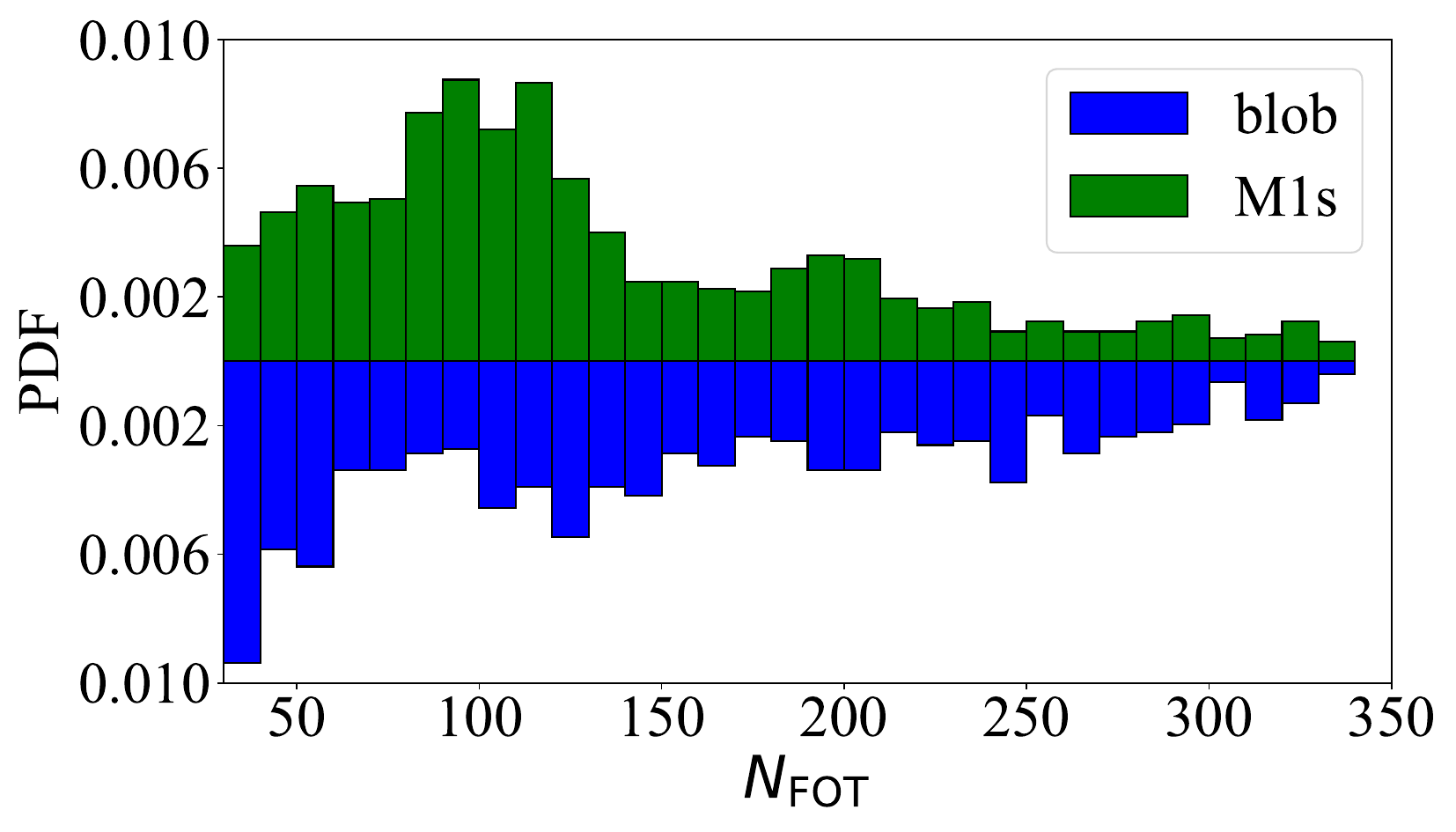}
	\caption{The probability density function diagram of $N_\textrm{FOT}$ of the tracking results of blob detection and M1s detection}
	\label{fig:PDF_tracking}
\end{figure}

To evaluate the improvement in tracking performance with the introduce of the machine learning methods, the probability density function (PDF) of the number of frames per track ($N_\textrm{FOT}$) based on blob detection and M1s within a single event are calculated, as shown in the Figure.~\ref{fig:PDF_tracking}.~Tracking results from M1s exhibits a higher probability density at around 100 frames per track.~In contrast, blob detection tracking shows a higher probability density around 50 frames per track.~After reviewing the tracking video and trajectories, it is observed that blob detection struggles with particles with soot.~As a result, the tracking using blob detection tends to pause at the soot formation area, resuming only after the soot disappears, thus continuing with the residuals of the combustion.

\subsection{Particle velocity statistic}

\begin{figure}[h!]
	\centering
	\includegraphics[width=70mm]{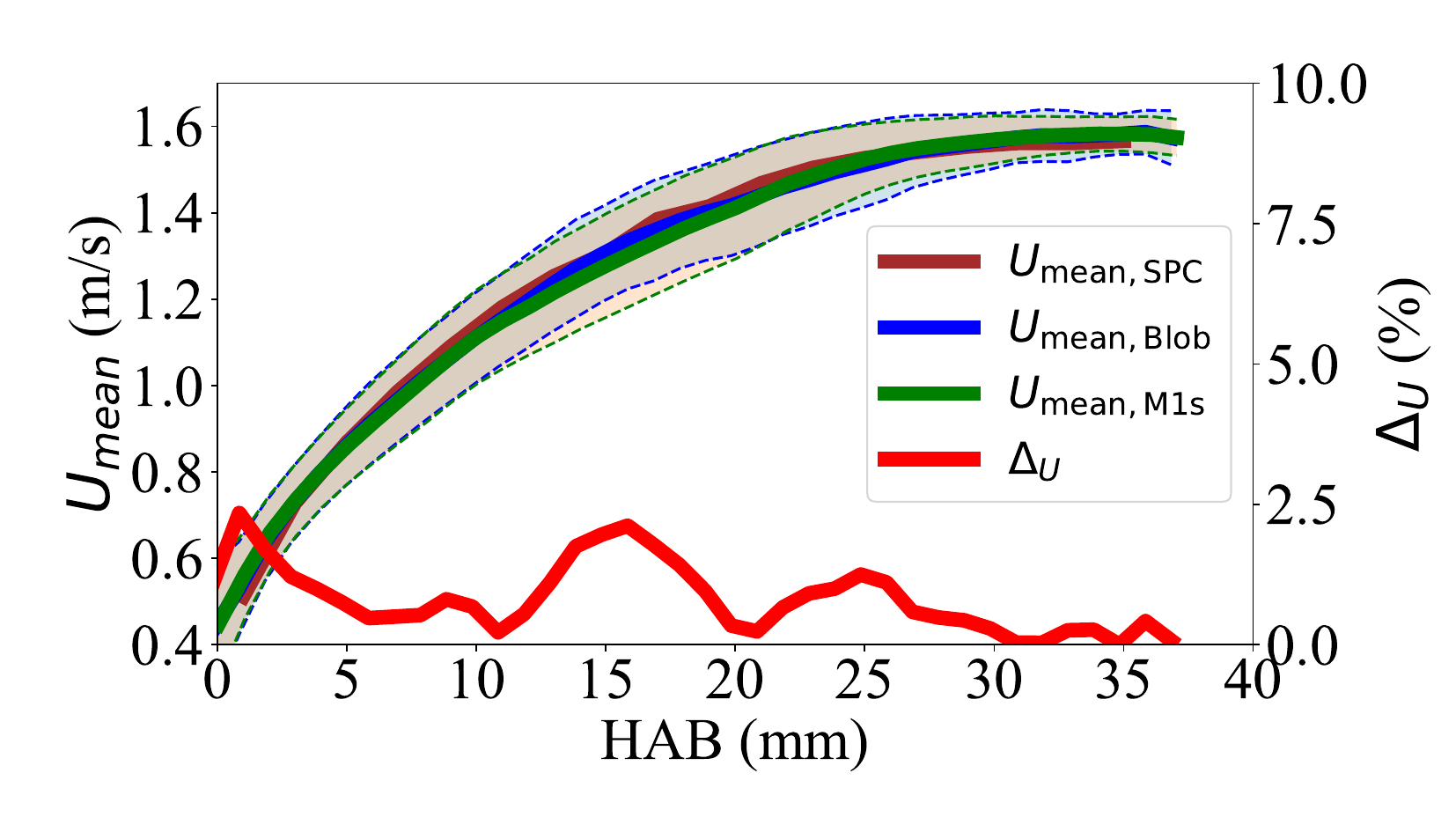}
	\caption{The averaged velocity and its standard deviation of case P040A20 using the tracking results of blob detection and M1s}
	\label{fig:v_mean_std}
\end{figure}

The motion behaviour of particles in low-PND scenarios are investigated by presenting the mean velocity ($U_\textrm{mean}$), its standard deviation, and the velocity difference of particles for the PND \textless\,0.18 ($N_\textrm{par}<$\,40) cases, derived from the tracking results of both blob detection and M1s detection, as shown in Figure.~\ref{fig:v_mean_std}.~The average velocities obtained from the two methods are similar, as shown the velocity difference ($\Delta U = \left| (U_\textrm{mean,Blob} - U_\textrm{mean,M1s}) \right|/ ((U_\textrm{mean,Blob} + U_\textrm{mean,M1s}) / 2) $) on the right y-axis.~Although blob detection-tracking does not sufficiently detect particles with soot, the average velocity is not influenced, since the SORT algorithm continues tracking particles in regions with less soot particles.~The standard deviation of particle velocity between HAB 10\,-\,25\,mm is higher than that at other heights, as this region corresponds to the primary area of soot formation.~The average velocity of the particles in single particle combustion (the brown line in the Figure), as denoted in previous work~\cite{Li.2022.Proc.Combust.Inst.}, is similar with the $U_{\textrm{mean}}$, since the particles in low-PND cases posses no interaction, which is similar with the single particle combustion cases.~As the PND is relatively low in this data set, particles eventually accelerate to match the ambient gas flow velocity, which is 1.67\,m/s for the A20 condition.

\begin{figure}[h!]
	\centering
	\includegraphics[width=\textwidth]{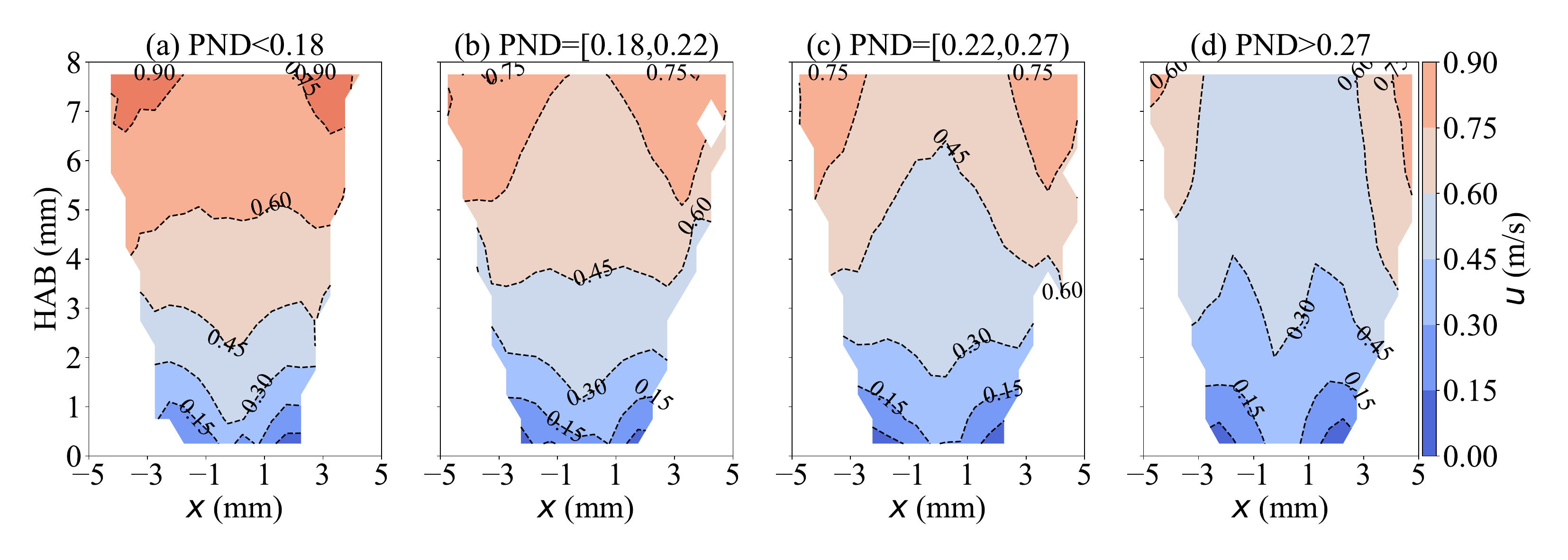}
	\caption{The axial velocity distribution contour of different PND cases}
	\label{fig:Vy_contour}
\end{figure}

To further investigate the particle-particle interaction of group particle combustion, the axial velocity ($u$) iso-contours of different PND cases are presented, as shown in Fig.~\ref{fig:Vy_contour}.~Tracked particles within each PND case in the HAB range 0\,-\,8 mm (before soot formation) were divided into 0.5\,mm grids, with the average velocity calculated for each grid cell to form the axial velocity iso-contour.~Generally, as PND increases, the particle jet expands outward, consistent with observations in Fig.~\ref{fig:V_jet}.~Additionally, the overall velocity decreases in higher PND cases.~At lower HAB (0\,-\,3 mm), particle velocity initially decreases before increasing in the radial direction.~However, at higher PND, the zone where velocity begins to increase shifts outward.~This trend aligns with findings from previous research~\cite{Li.2022.Proc.Combust.Inst.}, which demonstrated that the flame zone (responsible for accelerating particles through thermal expansion) tends to shift downward and outward with higher PND.~This shift of the flame zone location further explains the downward movement of the radial velocity increasing zone at higher HAB.



\section{Conclusions}
\label{sec:conclusion}
In this study, high-volatile bituminous coal combustion was conducted in a laminar flow reactor.~Particle positions were detected using the Mie-scattering imaging.~Traditional blob detection and machine learning detection methods, including YOLOv8 and RT-DETR, were applied to detect particles from different PND cases.~Particle tracking was subsequently performed using the SORT algorithm.~The reference volume was calculated to determine the PND.~Finally, the velocities of particles from different PND cases were analyzed.~The following conclusions can be drawn from the results:

(1) Machine learning methods demonstrated superior performance in particle detection.~Training the model on simple combustion cases and applied to predict complex cases proved feasible.~(2) YOLO combined with SAHI provided better detection on full-scale images, making it suitable for subsequent tracking.~(3) The SORT algorithm exhibited robust tracking capabilities for particles.~(4) The average velocity of particles in low PND cases was higher than in high PND cases, primarily due to increased particle interactions.

\section{Acknowledgement}
The authors kindly acknowledge financial support through Deutsche Forschungsgemeinschaft (DFG) - Projektnummer 215035359 - TRR 129 for its support through CRC/Transregio 129 "Oxy-flame: development of methods and models to describe solid fuel reactions within an oxy-fuel atmosphere."

\section{References}

{\bibliographystyle{mcs13}
\setlength{\bibsep}{0.5mm}
\def\section*#1{}
\bibliography{library}}

\end{document}